\documentstyle[12pt]{article}
\begin{document}

\begin{titlepage}
\begin{flushright}
FERMILAB-Pub-97/413-T\\
DTP/97/114\\
\end{flushright}
\begin{center}
{\Large\bf Probabilistic Jet Algorithms }\\
\vspace{1cm}
{\large
W.~T.~Giele$^1$ and E.~W.~N.~ Glover$^2$}\\
\vspace{0.5cm}
{\it
$^1$Fermi National Accelerator Laboratory, P.~O.~Box 500,
Batavia, IL 60510, U.S.A.} \\
{\it
$^2$Physics Department, University of Durham,  Durham DH1~3LE, England} \\
\vspace{0.5cm}
\end{center}
\begin{abstract}
Conventional jet algorithms are based on a deterministic view
of the underlying hard scattering process. Each outgoing parton
from the hard scattering is associated with a hard, well separated
jet. This approach is very successful because it allows
quantitative predictions using lowest order perturbation theory.
However, beyond leading
order in the coupling constant, when quantum fluctuations are included,
deterministic jet algorithms will become problematic
precisely
because they attempt
to describe an inherently stochastic quantum process using deterministic,
classical language.
This demands a
shift in the way we view jet algorithms.
We make a first
attempt at constructing more probabilistic jet algorithms that reflect
the properties of the underlying hard scattering
and explore the basic properties and problems of such an approach.
\end{abstract}
%%%%%%%PACS numbers: 12.38.Bx,  13.87.Ce,  12.38.Qk
\end{titlepage}

In high momentum
transfer scattering processes,  the concept  of jets makes a connection
between the hadron-level observations and the underlying partonic theory.
For ``good'' observables the theory is perturbatively calculable and,
at lowest
order (LO) in the coupling constant,  the predictions are
deterministic due to the absence of quantum fluctuations.
Each parton is associated with a high momentum jet.
After ``hadronizing'' the parton, one ends up with a collimated
shower of hadrons. Often color strings of hadrons between the partons
are introduced to model the energy flows better. However, the underlying
hard scattering in such an approach is still classical.
Conventional
jet algorithms are based on these models. Their main purpose is to
``invert'' the hadronization and identify the underlying hard scattering
parton structure. Within the classical approach this is perfectly
legitimate. In fact, jet algorithms are often compared by how well they
reconstruct the underlying partonic structure. Moreover, experimenters
use shower models such as HERWIG \cite{herwig}
to estimate their theory/experimental uncertainties by
hadronizing a parton in the detector simulation. Here the jet algorithm
is applied to estimate the mismeasurement of the original parton
energy and direction. One
then uses such models to either ``correct back'' to the parton level or
to absorb these effects into the systematic uncertainties.

This philosophy is acceptable as long as quantum fluctuations can
be neglected. The degree to which this approximation can be applied depends
on both the experimental accuracy and on the kinematics of the event (i.e.
well separated hard jets are, for all practical purposes, classical.).
However, when one counts jets using the classical jet algorithms, the majority
of the cross section in multi-jet events comes from the region where
the jet clusters are as close to each other as allowed by the jet algorithm
due to the collinear behaviour of QCD. This means that quantum corrections
are important as soon as the experimental uncertainties become small
enough. In recent years it has become apparent that one needs at least
next-to-leading order (NLO) perturbative calculations
to describe the precise jet data accumulated in current high energy scattering
experiments, i.e. one is sensitive to the quantum corrections. Applying
the usual deterministic jet algorithms
then leads to immediate problems which are
exactly associated with the stochastic nature of the
underlying scattering (i.e. each parton involved in the
the hard scattering is no longer identifiable as
a jet.).  Using shower models to correct back to the ``original'' parton
energies
might be misleading.
Furthermore, adding some arbitrary soft radiation or
collinear fragmentation can significantly alter the jet energies and directions
found using  deterministic jet algorithms
rendering the theoretical predictions infrared unstable.
Similarly, in the experiment
a single hit in the calorimeter should only be associated with a jet/cluster
in a probabilistic manner since small mismeasurements will change the
assignment
of calorimeter hits to jets.
These instabilities are reflected
in large hadronization corrections and large experimental uncertainties.

The obvious way out is to introduce probabilistic jet algorithms to reflect
the quantum mechanical nature of the hard scattering. On an
{\it event by event} basis the algorithm will give a probability distribution
to observables in the event (e.g. transverse
energy, number of jets, jet-jet mass, etc.). This is contrary to classical
jet algorithms which give a definite value to the observables. There will
be an immediate impact on the hadronization effects and measurement errors. For
example, the number of jets found using deterministic jet algorithms will
always be
an integer number.
This number can vary on an event by event basis due to fluctuations
in the hadronization process and the measurement errors (even
if the underlying hard scattering is kept fixed). However with a
probabilistic jet algorithm, each jet topology is associated with a probability
(or equivalently the event contains a fractional number of jets).
Hadronization and
measurement uncertainties will still alter the probability of finding a given
jet multiplicity, but now will be far less important, reflecting
far more accurately the properties of the underlying quantum process.
This will have some influence on observables that depend
on the number of jets in the event such as the
identification of top quark events.
Any quantum algorithm must revert back to the classical algorithm  in the
limit of well separated, high energy jets (i.e. one of the jet multiplicities
has a probability approaching 100\%.).

Recently, Tkachov \cite{FVT} has used event shape variables that satisfy
calorimetric continuity (jet discriminators) to describe the jet topology
of the event.
Any event shape observable such as the C-parameter \cite{C}, that vanishes
in the two-jet limit may be considered a three-jet discriminator.
These multi particle correlators are continuous and are also stable against
small variations of the input.

In the rest of the paper we will define and explore the above
concepts in some detail. Also we will give an explicit example
of an observable calculated with a probabilistic algorithm. This
example is taken from hadron colliders: the triple differential
di-jet cross section.

In order to test our ideas and to examine how hadronisation corrections
influence
the counting of the number of jets, we  consider a simple one-dimensional
toy model with two partons of energies $E_1$ and $E_2$, separated in
azimuthal angle, $\phi$, by $\Delta R$, recoiling against a colour
neutral object at $\phi = \pi$.
Typically, we choose $E_1 = 100$~GeV and $E_2= 50$~GeV.
Although the formation of jets should not depend on the details of
hadronisation, it is important to have a semi-realistic model
of how the partonic energy flow fragments into hadrons.
The simplest model of hadronisation is Feynman's ``tube'' model \cite{tube}
where a parton produces a jet of light hadrons uniformly
distributed in rapidity along the jet direction.
Light hadrons are distributed in transverse momentum $p_T$ (again with
respect to the jet axis) with a gaussian
density $\rho(p_T)$ such that the average transverse momentum
is controlled by the parameter, $\lambda$, that sets the hadronisation scale,
$\langle p_T\rangle = \lambda$.
Reasonable choices of $\lambda$ lie in the range
$\lambda \sim 0.5 - 1$~GeV.
We use a Monte Carlo approach to fragment the parton into several
hadrons which obey the transverse momentum constraint,
$\sum_{i} \vec{p}_{Ti}  = \vec{0}$, where the sum runs over the
hadrons produced by the parton.
Each individual parton ``shower'' produces ${\cal O}(10)$ hadrons.
This is repeated many times for each parton configuration to generate
a sample of ``data''.

We examine two distinct jet algorithms that are representative of
those commonly used in high energy hadron collider experiments.
In each case we use a cone size $R=0.7$.\\
(a) {\em The fixed cone \cite{UA2alg}.}  Place the cone around the
highest energy particle not already in a jet and add up the energy

inside the cone.
If $E> E_{\rm cut}$, it is a jet.  Repeat.\\
(b) {\em The iterative cone \cite{CDFalg,D0alg}.}  Place the cone
around the highest energy particle not already in a jet.   Find the
energy weighted axis and move the cone to that location.
Repeat until jet stabilises.
Repeat until no more jets are found, then resolve overlaps - i.e.
for particles that are in more than one jet distribute energy amongst
jets or merge clusters.
Add up the energy inside the cluster and count jets with $E> E_{\rm cut}$.\\

At tree level, both algorithms give the same prediction for the number of
jets.  If the spacing between partons is less than the cone size,
$\Delta R < R$, then the partons coalesce and only one jet is found.
For all
other values of $\Delta R$ two jets are observed.
Once hadronisation is switched on (i.e. $\lambda = 1$~GeV), the number of
jets observed depends on exactly how the hadrons are distributed in azimuth.
Generally, for $\Delta R \sim 0$ and $\Delta R \sim 2R$, either one or two
jets
with $E > 30$~GeV are found. This is not always the case and sometimes the
hadronisation pattern will cause additional jets to be found.
However, for fixed intermediate values of $\Delta R$, and particularly those
values close to $R$, consecutive events will flip between finding one or
two jets.   Depending on the actual distribution of hadrons, the topology
of the event undergoes a catastrophic change.  This is reflected in a
large variance on the average number of jets found in the event after
averaging over 10000 events.
Figs. 1(a) and 2(a) illustrate this for each algorithm.
By the nature of the algorithm, an integer number of jets is found in
each event, but the average number (for $\Delta R \sim R$) is close to
1.5 (with a standard deviation of close to 0.5).

\begin{figure}[t]\vspace{8cm}
\includegraphics{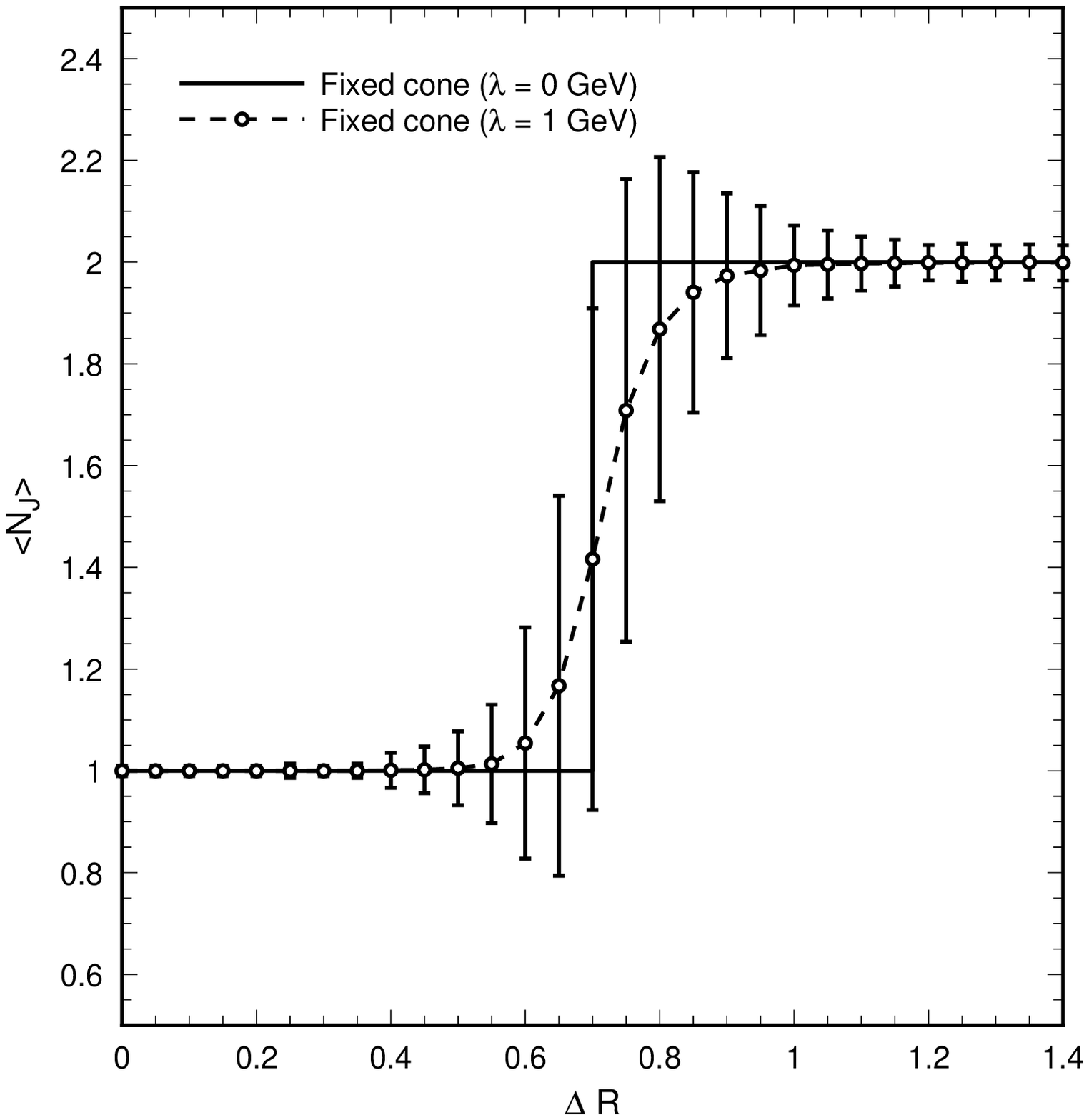}
\includegraphics{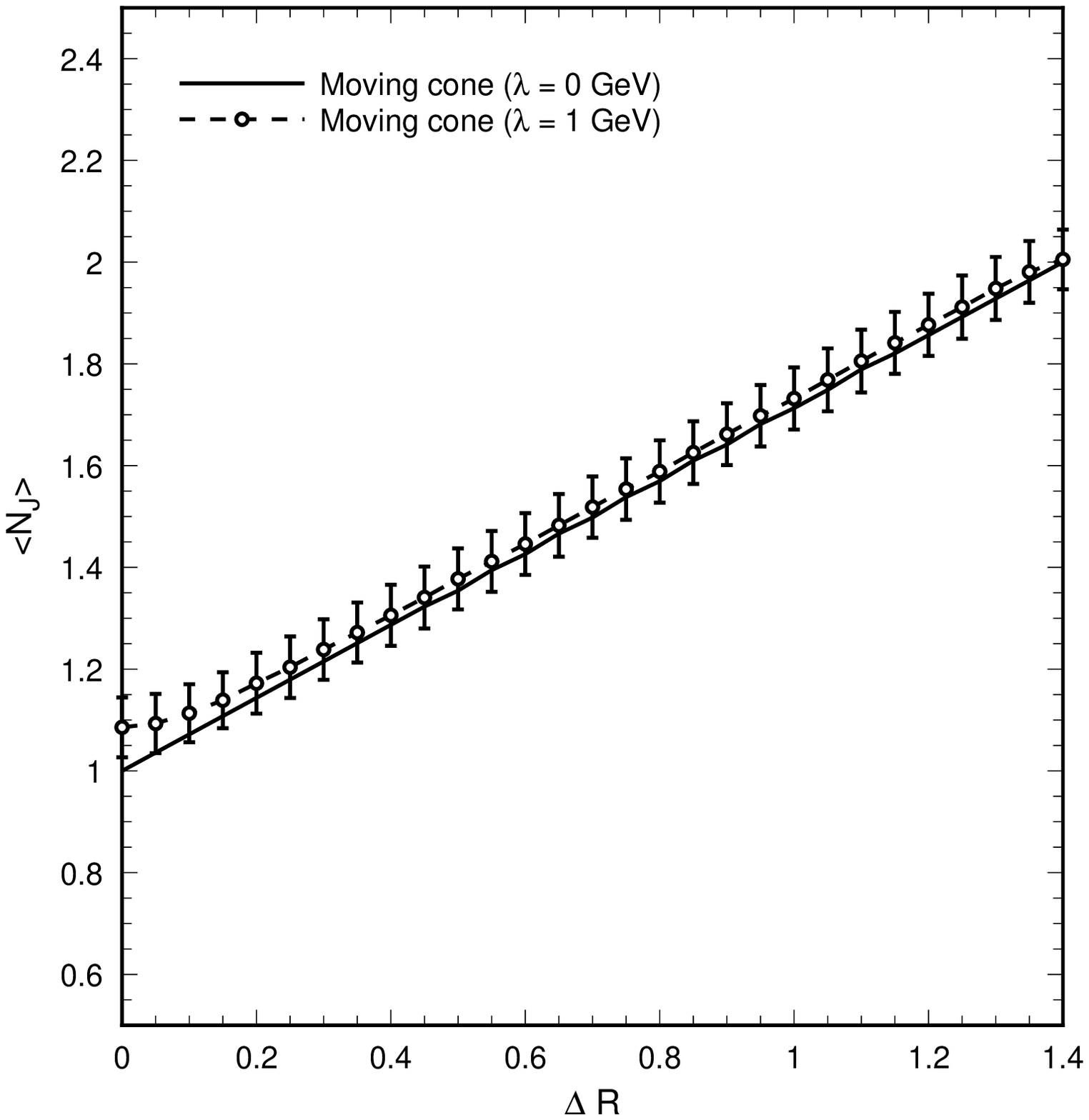}
\caption[]{The average number of jets observed as a function of the
parton-parton separation $\Delta R$ using the
(a) fixed cone (b) moving fixed cone.   The parton level results
($\lambda = 0$) are shown as solid lines, while the hadron level
($\lambda=1$) predictions are shown as
a dashed line joining data points with standard deviation obtained from
a large number of hadronised events.}
\end{figure}

Varying the hadronisation scale $\lambda$ alters the details, but does not
alter the gross feature: The individual event is not representative of
the average. The problem can be traced back to the way in which the jet
algorithm is applied.  In each case, the starting point is the most energetic
particle (or the least energetic in the case of the $K_T$ algorithm \cite{KT}).
This choice is very sensitive to collinear fragmentation, soft radiation,
the details
of the hadronisation and also possible mismeasurement of the particle
energy by the detector.

\begin{figure}[t]\vspace{8cm}
\includegraphics{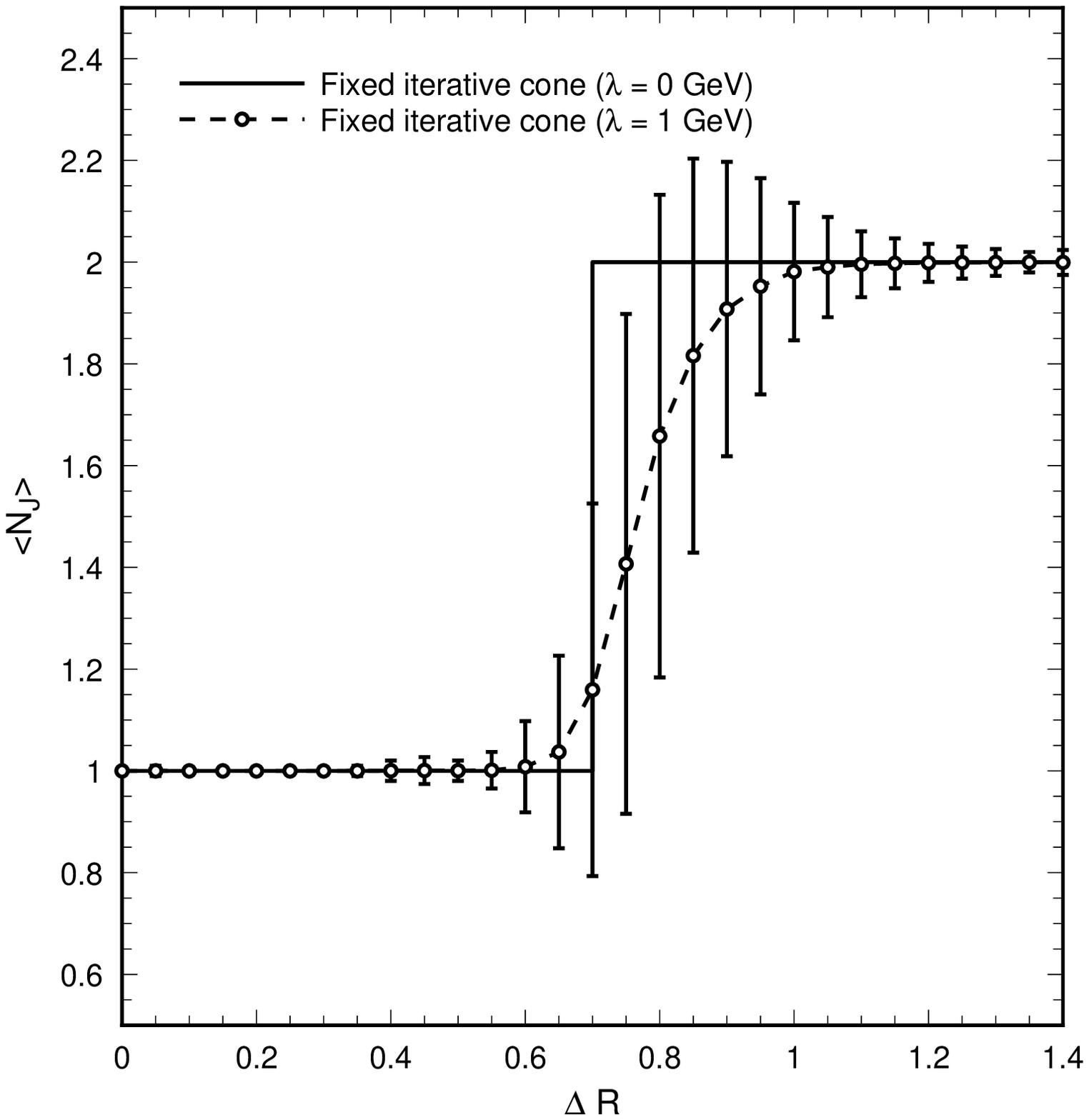}
\includegraphics{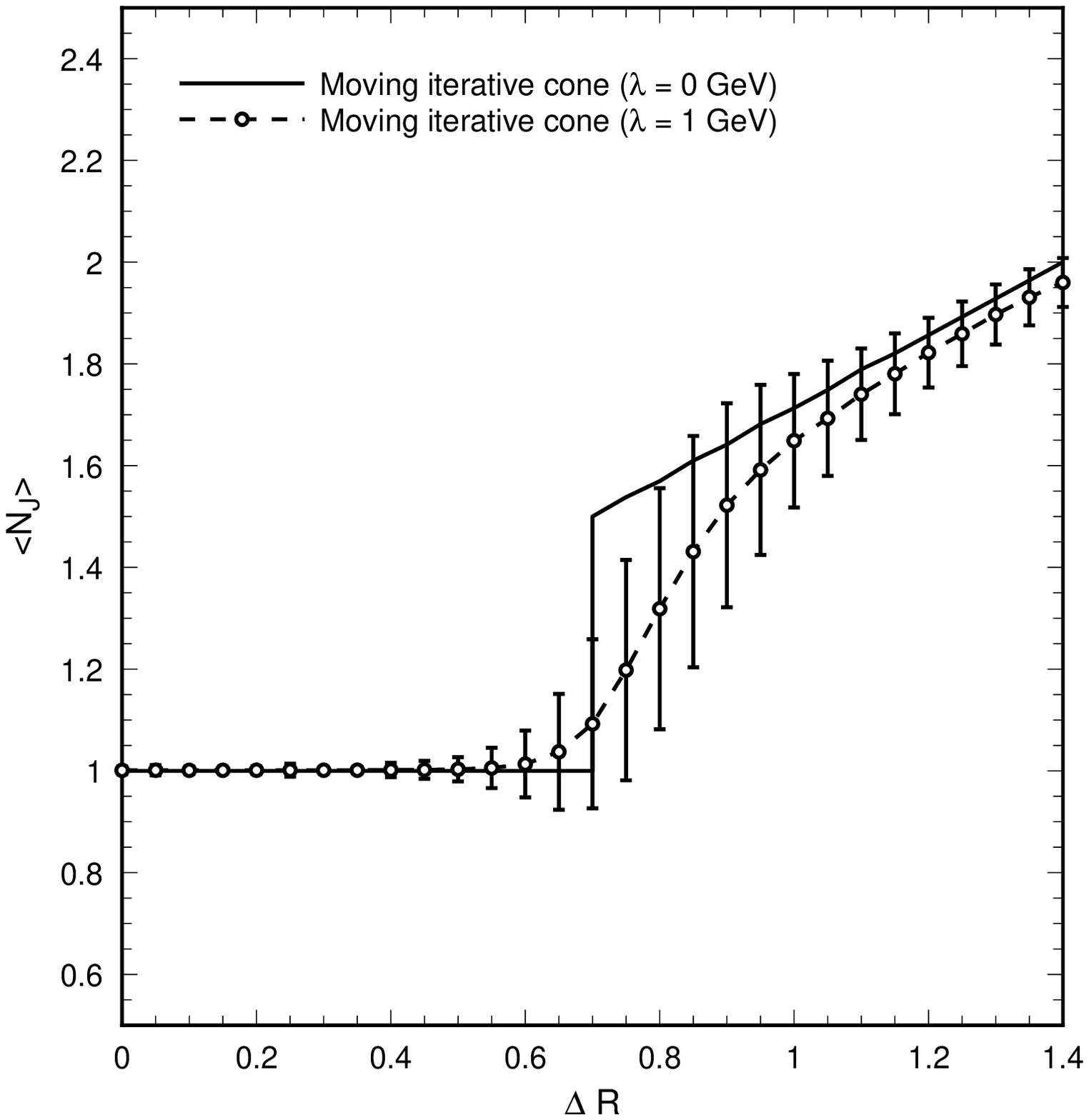}
\caption[]{The average number of jets observed as a function of the
parton-parton separation $\Delta R$ using the
(a) iterative cone (b) moving iterative cone.   The parton level results
($\lambda = 0$) are shown as solid lines, while the hadron level
($\lambda=1$) predictions are shown as
a dashed line joining data points with standard deviation obtained from
a large number of hadronised events.}
\end{figure}

Any algorithm that requires an integer number of jets to be found in the event
will have the same problems.
As discussed earlier, an approach in which a probability is associated with
each
jet topology is more natural.
We can easily adapt the existing algorithms to the probabilistic approach by
treating each hadron or calorimeter cell as the seed tower for the algorithm.
Each starting point will generate a particular jet configuration, and
by averaging over configurations we can obtain probabilities of finding
that particular jet.
Because we now consider a varying starting point for each jet, we denote
such algorithms as {\em moving}.   The analogues of the two earlier jet
algorithms
are:\\
(A) {\em The moving fixed cone.}  Center the cone on a calorimeter cell
and add up the energy inside the cone.
If $E> E_{\rm cut}$, it is a jet.  Repeat. Repeat for all calorimeter cells.\\
(B) {\em The moving iterative cone.}  Center the cone on a calorimeter cell.
Find the energy weighted axis and move the cone to that location.
Repeat until jet stabilises.
If $E> E_{\rm cut}$ {\em and} the starting calorimeter cell lies within the
cone, it is a jet. Repeat for all calorimeter cells.
Note that even in the case of multiple clusters, there is never the issue of
overlapping cones. Each calorimeter cell contributes fully to each cone
containing it.\\
Similar probabilistic algorithms could be constructed for the $K_T$ or
Durham algorithm \cite{KT,durham}.

In each case, each starting point (and each resulting jet energy) obtains
a weight of $1/2 R$ to normalise the probability of finding a jet.
To compute the number of jets, we count the number of calorimeter cells that
give rise to an allowable jet and divide by the number of cells in
$2R$ of azimuth.
Even at the parton level, the algorithms behave slightly differently.
For $\Delta R < R$,   each calorimeter cell within
$R$ of the  final jet axis reconstructs the energy and direction of
the parton for
algorithm (B)  and a single jet is found with energy $E_1+E_2$.
However, for the moving cone (algorithm (A)), sometimes the cone contains
one parton and sometimes two.  That is to say, sometimes we find two jets,
with energies $E_1$ and $E_2$ and sometimes a single jet with energy
$E_1+E_2$.
For $2R > \Delta R > R$, both algorithms find a combination of one-jet
and two-jet configurations - calorimeter cells between the partons are
more likely to result in one-jet configurations while those outside
the partons will
find single parton jets.
For well separated partons, $\Delta R > 2R$, then we always find two
jets with energies $E_1$ and $E_2$.   This is as it ought to be,
for events with two well separated clusters should be
classified as two jet events with a very high probability.
Only in the intermediate regions should the one or two jet topologies
have probabilities far from either 0\% or 100\%.
At $\Delta R = R$, we see that each algorithm gives the average number
of jets as 1.5.   Note that this is obtained with a single event and
the individual event is now representative of the average.
While hadronisation will influence the formation of jets and will alter the
one and two-jet probabilities, there will no longer be the catastrophic
swapping between topologies.
This can be seen in Figs. 1(b) and 2(b) where the average number of jets
produced in the same data sample of hadronised events used earlier
is plotted against the parton-parton separation, $\Delta R$.
In all cases, the fluctuations are significantly smaller compared
to those obtained with the corresponding conventional deterministic algorithm.
Varying the hadronisation scale $\lambda$ and the jet energy cut $E_{\rm cut}$
changes the details of the plots, but in all cases, the probabilistic
algorithms give smaller variances on $\langle N_J\rangle$ than the
deterministic ones.
We also see that the moving fixed cone algorithm (A) is rather good
for counting clusters with an energy bigger than some threshold, but
that the moving iterative cone algorithm (B) has a good jet energy resolution.

\begin{figure}[t]\vspace{8cm}
\includegraphics{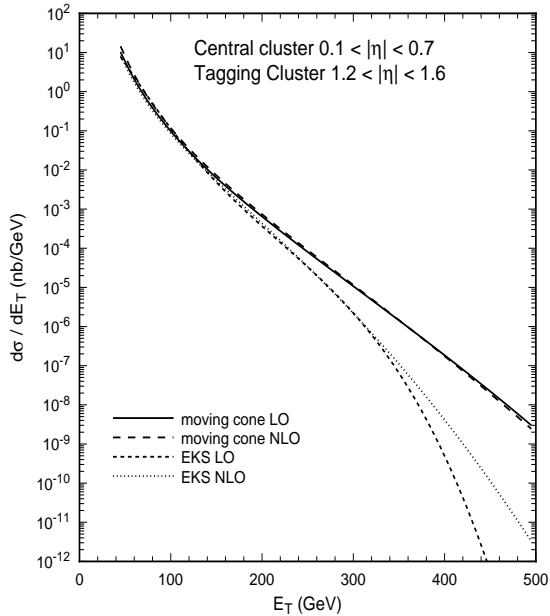}
\caption[]{The leading (LO) and next-to-leading order (NLO) predictions
using both the moving fixed cone and the EKS jet algorithm \cite{EKS} for
$\int d\eta_1 d\eta_2 d\sigma /dE_Td\eta_1d\eta_2  $ for
$0.1 < |\eta_1| < 0.7$ and $1.2 < |\eta_2| < 1.6$.
We use CTEQ4M parton distributions \cite{CTEQ4} with
renormalisation/factorisation scale $\mu = \sum E_T /2$.}
\end{figure}

As an example of how probabilistic algorithms may be applied to calculate
more realistic quantities, we consider the triple differential dijet
distribution.   Here, one jet with transverse energy $E_{T1}$ is produced
in a central rapidity strip $0.1 < |\eta_1| < 0.7$, with a second tagging
jet with $E_{T2} > E_{T1}/2$
in a more forward rapidity slice $1.2 < |\eta_2| < 1.6$.
This observable has been studied both experimentally \cite{CDF} and
theoretically \cite{dijet}.
We use parton level Monte Carlo JETRAD \cite{dijet,jetrad}
with the moving fixed cone algorithm (A) and allow one cone of radius $R=0.7$
to move in the central region and one in the forward region.
The rapidity cuts are applied to the
cone axis. Note that this extends the range of the allowed parton rapidity in
a probabilistic manner.
 For each cone position, the partonic energy inside the cone is
calculated and entered in the distribution (with weight $1/\pi R^2$).
The results are shown in Fig.~3, where we also show the distribution obtained
using the EKS jet algorithm \cite{EKS} (and $R_{sep} = 1.3$ \cite{Rsep}).
Events are triggered by requiring a summed $E_T$ of at least 80~GeV is
registered in the calorimeter.

We could, for instance, also construct more involved correlators such
as the two-cluster mass using one of the moving cone algorithms. Here
there would be two cones moving over the detector and one would compute
the mass of the pair.
It is important to note that the cones are {\em not} mutually exclusive.
In fact, there will be self correlations between overlapping cones.
However, particles in the overlap region are assigned to both cones
and there is never any question of which particle goes to which cone.

We have demonstrated that the idea of probabilistic jet algorithms
can be applied to jet events in a sensible manner. At all stages of
the event evolution (hard parton scattering, hadronization  and
measurement) the probabilistic jet algorithm will give a probability
to the observable that is close to the average on an
{\it event by event} basis. This is in contrast to the deterministic
algorithms, which on an event by event basis will fluctuate significantly.
As the underlying quantum mechanical hard scattering is by definition
stochastic these results should not come as a surprise.

EWNG thanks the Fermilab Theory Group for their kind hospitality
during the period in which this research was carried out.

\end{document}